\documentclass[%
 reprint,
showpacs,preprintnumbers,
 amsmath,amssymb,
 prl,
]{revtex4-1}

\usepackage{graphicx}
\usepackage{dcolumn}
\usepackage{bm}

\begin{document}

\preprint{APS/123-QED}

\title{Spin-gap opening accompanied by a strong magnetoelastic response in the \textit{S} = 1 magnetic dimer system Ba$_3$BiRu$_2$O$_9$}

\author{Wojciech Miiller,$^{1,2}$ Maxim Avdeev,$^{2}$ Qingdi Zhou,$^{1}$ Andrew J. Studer,$^{2}$ Brendan J. Kennedy,$^{1}$ Gordon J. Kearley,$^{2}$ and Chris D. Ling$^{1}$}

\affiliation{$^{1}$School of Chemistry, The University of Sydney, Sydney, NSW 2006, Australia}
 
\affiliation{$^{2}$The Bragg Institute, ANSTO, PMB 1, Menai 2234, Australia}
 
\date{\today}

\begin{abstract}

Neutron diffraction, magnetization, resistivity, and heat capacity measurements on the 6H-perovskite Ba$_{3}$BiRu$_2$O$_9$ reveal simultaneous magnetic and structural dimerization driven by strong magnetoelastic coupling. An isostructural but strongly displacive first-order transition on cooling through $T^*=176$ K is associated with a change in the nature of direct Ru$-$Ru bonds within Ru$_{2}$O$_{9}$ face-sharing octahedra. Above $T^*$, Ba$_{3}$BiRu$_2$O$_9$ is an $S=1$ magnetic dimer system with intradimer exchange interactions $J_0/k_{\rm{B}}=320$ K and interdimer exchange interactions $J'/k_{\rm{B}}=-160$ K. Below $T^*$, a spin-gapped state emerges with $\Delta\approx220$ K. \textit{Ab initio} calculations confirm antiferromagnetic exchange within dimers, but the transition is not accompanied by long range-magnetic order.

\end{abstract}

\maketitle

Quantum cooperative phenomena involving charge, spin, orbital, and lattice order parameters are among the most explored areas of modern solid-state physics. A particularly rich vein of quantum cooperative phenomena are low-dimensional magnetic systems featuring motifs such as dimers, chains, ladders, and plaquettes, where coupling between spin and lattice degrees of freedom can give rise to magnetoelastic effects such as the spin-Peierls transition in $S=1/2$ Ising-chain antiferromagnets \cite{PhysRevLett.70.3651}. The most thoroughly studied are based on $S=1/2$ 3\textit{d} transition metal cations such as Cu$^{2+}$, V$^{4+}$ and Ti$^{3+}$. The less common $S=1$ systems remain relatively neglected, although $S=1$ ruthenates (low-spin Ru$^{4+}$), in particular, show fascinating properties including superconductivity (Sr$_2$RuO$_4$\cite{1998Natur.394..558L}), non-Fermi liquid behaviour (La$_4$Ru$_6$O$_{19}$\cite{2001Natur.411..669K}) and low-dimensional character (the famous spin-gapped Haldane phase Tl$_2$Ru$_2$O$_7$\cite{2006NatMa...5..471L}).

\begin{figure}
\includegraphics[width=5.17cm, angle=270]{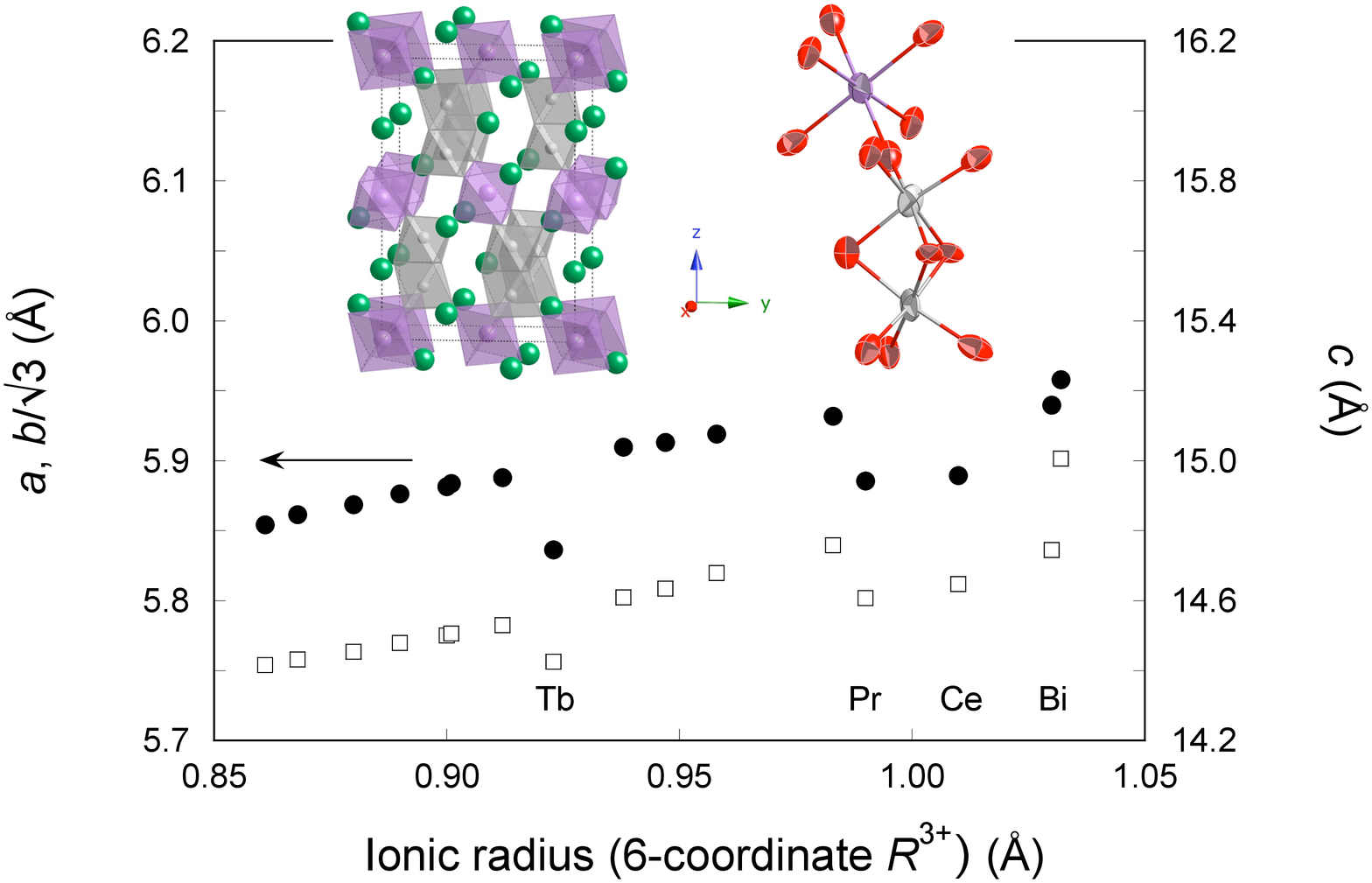}%
\caption{\label{fig1}(Color online) Ionic radii \cite{1976AcCrA..32..751S} of $R^{3+}$ ($R$ = Bi, Y, rare-earth) in 6-fold coordination $vs.$ unit cell lengths $a$, $b/\sqrt{3}$ (filled circles) and $c$ (open squares) for Ba$_{3}R$Ru$_{2}$O$_{9}$ at 298 K. Error bars are smaller than symbols. Adapted from Doi \textit{et al.} \cite{B105134M} and references therein, with $R$ = Bi from the present work. The refined (HRPD@ISIS) 2 K structure of Ba$_3$BiRu$_2$O$_9$ is shown with BiO$_6$ octahedra (\textit{violet}), Ru$_2$O$_9$ dimers (\textit{grey}), Ba$^{2+}$ (\textit{green}, and O$^{2-}$ ions (\textit{red}). Polyhedral units are expanded and drawn with 90\% probability thermal ellipsoids.}
\end{figure}

This report concerns Ba$_3$BiRu$_2$O$_9$, first reported by Darriet \textit{et al.}\cite{Darriet} as a 6H-perovskite with a small monoclinic (\textit{C}2/\textit{c}) distortion. It contains BiO$_6$ octahedra sharing vertices with Ru$_2$O$_9$ face-sharing octahedral dimers, and Ba$^{2+}$ ions occupying high-coordinate \textit{A} sites (Fig. \ref{fig1}). It forms part of the Ba$_3$\textit{R}Ru$_2$O$_9$ series, where \textit{R} is a rare-earth or 3\textit{d} transition metal\cite{B105134M,B111504A}, indium\cite{PhysRevB.58.10315} or zirconium\cite{ZAAC:ZAAC200500442}. The rare-earth $R$ cations usually have a 3+ oxidation state, giving Ru$_{2}^{9+}$ dimers; but Tb, Pr, and Ce have 4+ oxidation states, giving Ru$_{2}^{8+}$ dimers, and shrinking the unit cell due to the reduced ionic radius of $R$ \cite{1976AcCrA..32..751S} (Fig. \ref{fig1}). Our experimental lattice parameters for Ba$_{3}$BiRu$_{2}$O$_{9}$ strongly suggest that it belongs in this Ru$_{2}^{8+}$ category. 

Polycrystalline Ba$_3$BiRu$_2$O$_9$ was synthesized by solid-state reaction. Neutron powder diffraction (NPD) data were collected on the instrument Wombat at OPAL, ANSTO, from $2-300$ K using 2.9609 {\AA} neutrons; and at 2 K on the HRPD diffractometer at ISIS. Structure refinements \textit{via} the Rietveld method were carried out using GSAS\cite{citeulike:9434647} with the EXPGUI front-end\cite{EXPGUI}. Magnetic susceptibility, electrical resistivity and heat capacity were measured in a Quantum Design PPMS. \textit{Ab initio} calculations were performed in the generalized gradient approximation (GGA) using the Vienna \textit{ab initio} Simulations Package (VASP 5.2)\cite{PhysRevB.54.11169}. A supercell containing two primitive cells (60 atoms) was used with standard PAW potentials\cite{PhysRevB.59.1758}, a \textit{k}-mesh of 162 points in the irreducible Brillouin zone wedge and a cutoff energy of 450 eV. Total energy converged to within 10$^{-5}$ eV.

\begin{figure}
\includegraphics[width=8.5cm]{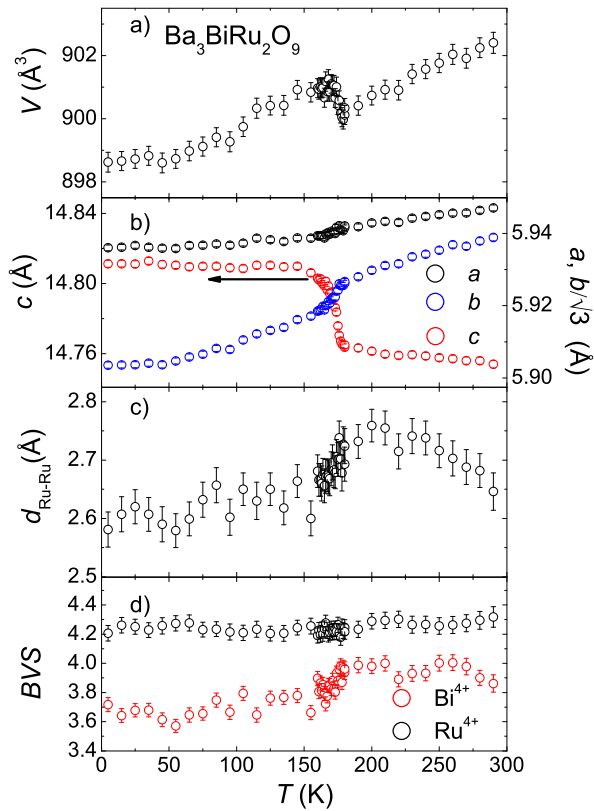}%
\caption{\label{fig2}(Color online) Temperature dependencies of a) unit cell volume $V$, b) unit cell lengths, c) intradimer $d_{\rm{Ru-Ru}}$ distance, and d) bond valence sums  \cite{Brese:st0462} for Bi and Ru from Rietveld-refinement against NPD (Wombat@OPAL) data. }
\end{figure}

Fig. \ref{fig2} displays the temperature dependencies of the unit cell volume, lattice parameters, and intradimer $d_{\rm{Ru-Ru}}$ distance. The volume shows normal thermal contraction down to $\sim$180 K, then increases rapidly, mostly by elongation of the stacking \textit{c} axis (Fig. \ref{fig2}b). The transition appears to be first order in nature, with the slightly rounded disjunctions in refined structural parameters being ascribed to the presence of both high- and low-temperature forms close to the transition (note that only a single phase could be refined at all temperatures, due to the resolution of the diffractometer used). Interestingly, the intradimer distance $d_{\rm{Ru-Ru}}$ $-$ which lies along \textit{c} $-$ shows the opposite behaviour, shrinking from a maximum 2.76(3) {\AA} at 200 K to 2.66(3) {\AA} at 165 K. 

The apparent 4+ oxidation state for the single crystallographically unique Bi site in Ba$_3$BiRu$_2$O$_9$ is very unusual (it would have an unstable $5d^{10}6s^{1}$ valence configuration), leading us to consider whether the structural transition is driven by disproportionation (2Bi$^{4+}$ $\rightarrow$ Bi$^{3+}$ + Bi$^{5+}$) or charge transfer (Bi$^{4+}$ + 2Ru$^{4+}$ $\rightarrow$ Bi$^{3+}$ + 2Ru$^{4.5+}$). Either process should be obvious in bond valence sums (BVS) \cite{Brese:st0462} calculated from experimentally refined structure parameters. BVS (Fig. \ref{fig2}d) show no anomaly for Ru on cooling, ruling out charge transfer. For Bi they shows a small anomaly; however, our NPD data show no evidence for symmetry lowering (additional or split peaks below the transition) due to long-range Bi$^{3+}$:Bi$^{5+}$ order. Moreover, Bi$^{4+}$ oxides generally disproportionate at much higher temperatures than the transition of interest here (\textit{e.g.}, two distinct Bi sites can be distinguished by BVS in BaBiO$_{6}$ even at 900 K \cite{Kennedy:ws5040}). The $6s$ electron on Bi$^{4+}$ probably localises at similarly high temperatures in Ba$_3$BiRu$_2$O$_9$, but long-range Bi$^{3+}$:Bi$^{5+}$ order is frustrated by the triangular disposition of Bi sites, with the only evidence being slightly anisotropic oxygen atomic displacement parameters (ADPs) (Fig. \ref{fig1}). Most importantly, ADPs and diffraction peak widths show no discontinuities at the structural transition to indicate a significant change in local disorder or lattice strain. 

It does seem likely that \textquoteleft\textquoteleft Bi$^{4+}$\textquoteright\textquoteright\space plays an important role in the transition at $T^*$, considering that no such transition is observed for any $R^{4+}$ ($R$ = rare earth) cation. Unfortunately, the evidence at this stage is insufficient to identify the precise nature of that role. One possibility is that Bi$^{4+}$ may be acting as a transient charge reservoir, communicating changes in the electronic state between isolated Ru$_2$O$_9$ dimers as they switch from the high- and low-temperature forms, thereby facilitating a first-order transition that never takes place where more stable $R^{4+}$ cations are involved.

\begin{figure}
\includegraphics[width=8.5cm]{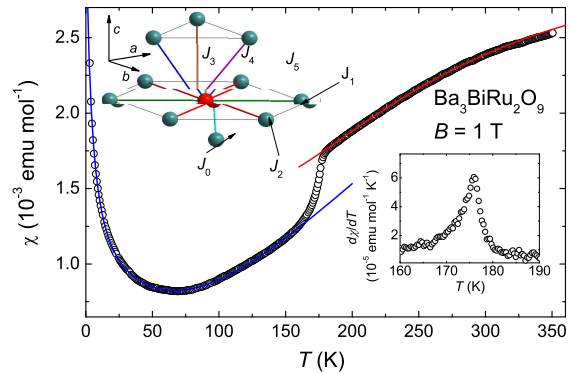}%
\caption{\label{fig3}(Color online) Magnetic susceptibility measurements. The solid lines are fits: eq. \ref{eq.2} - \textit{red} line and eq. \ref{eq.3} - \textit{blue} line. The bottom right inset shows the temperature dependence of the $\rm{d}\chi/\rm{d}T$ derivative. The upper left inset displays the magnetic exchange interactions with nearest ($J_0$ - within a dimer) and next-nearest ($J_i$, $i=1...5$) neighbors.}
\end{figure}

Molar magnetic susceptibility $\chi$ is shown in Fig. \ref{fig3}. The structure consists of effectively isolated Ru$_2$O$_9$ units, so the expression for isolated $S=1$ dimers:

\begin{equation}
\chi_{\rm{iso}}=\frac{2N_{\rm{A}}g^2\mu_{\rm{B}}^2(1+5\exp(-2J_0/k_{\rm{B}}T))}{k_{\rm{B}}T(3+\exp(J_0/k_{\rm{B}}T)+5\exp(-2J_0/k_{\rm{B}}T))}
\label{eq.1}
\end{equation}

should be appropriate, where N$_{\rm{A}}$ is Avogadro's number, $\mu_{\rm{B}}$ is a Bohr magneton, $k_{\rm{B}}$ is the Boltzmann constant, $g$ is the standard electron $g=2$ factor and $J_0$ is the intradimer magnetic coupling. However, a much better quality of fit is obtained if interdimer interactions $J'$ are included in the calculations in the mean-field approximation, so susceptibility $\chi_{\rm{dim}}$ can be written as:

\begin{equation}
\chi_{\rm{dim}}=\frac{\chi_{\rm{iso}}}{(1+\lambda\chi_{\rm{iso}})}
\label{eq.2} 
\end{equation}

where $\lambda = J'/N_{\rm{A}}g^2\mu_{\rm{B}}^2$. A fit above 180 K $\chi(T)=(1-x)\chi_{\rm{dim}}+\chi_0+xC/(T-\Theta)$, which combines dimer, temperature independent and Curie-Weiss contributions ($C=1$ emu K mole$^{-1}$ for $S=1$ impurities) respectively, is shown. The best fit is to $J_0/k_{\rm{B}}=320(20)$ K, $J'/k_{\rm{B}}=-160(10)$ K, $x=0.02$, $\Theta=0$ K and $\chi_0=0.0011$ emu mole$^{-1}$. This is comparable to Ba$_3$PrRu$_2$O$_9$\cite{B105134M}, where $J_0/k_{\rm{B}}=$280 K. A comment should be made here concerning next-nearest neighbor interactions. While nearest neighbors are obviously those within Ru$_2$O$_9$ units, higher-order neighbors are less obvious. Each Ru$^{4+}$ is surrounded by 6 others in the pseudo-hexagonal $ab$ plane and 3 from the neighboring plane, connected \textit{via} Ru-O-Bi-O-Ru superexchange paths of approximately equal length. In the $C2/c$ space group, these 9 sites comprise 5 inequivalent positions, leading to 5 different exchange constants $J_i$, $i=1...5$ (inset in Fig. \ref{fig3}). As $J'=\sum_{i=1}^{N=5}n_iJ_i$, where $n_i$ is the multiplicity of next-nearest neighbors interacting \textit{via} $J_i$, the average interdimer exchange parameter is $\overline{J}_{i}/k_{\rm{B}}=-17.7$ K. The intradimer coupling $J_0$ is about double the absolute value of this interdimer $J'$. Susceptibility results thus show that above $T^*=176$ K, Ba$_3$BiRu$_2$O$_9$ is an antiferromagnetic (AFM) dimer system with weak ferromagnetic (FM) next-nearest neighbours interactions.

A drastic drop in magnetic susceptibility below $T^*$ is highlighted by a maximum in the temperature derivative of $\chi(T)$ (inset in Fig. \ref{fig3}). As $T^*$ coincides with the discontinuities in cell volume and $d_{Ru-Ru}$, Ba$_3$BiRu$_2$O$_9$ appears to undergo coincident structural/magnetic dimerization. We tentatively ascribe this to the opening of a gap in the spin-excitation spectrum between non-magnetic singlet ground state and excited triplet spin configurations within dimers. The magnetic susceptibility of such a system should follow \cite{2001Natur.411..669K}:

\begin{equation}
\chi_{\rm{sg}} = aT^{0.5}\exp(\frac{-\Delta_{\chi}}{k_{\rm{B}}T})
\label{eq.3}
\end{equation}

where $a$ is a constant and $\Delta_{\chi}$ is the value of the spin gap. A least-square fit performed in 2-150 K temperature range to the equation: $\chi(T)=\chi_{\rm{sg}}+\chi_0+c/(T-\Theta)$ (shown in Fig. \ref{fig3}, \textit{blue} line) yields $a=0.00021$ emu mol$^{-1}$K$^{-0.5}$, $\Delta_{\chi}/k_{\rm{B}}=247(2)$ K, $\chi_0=0.0006$ emu mol$^{-1}$, $c=0.013$ emu mol$^{-1}$K$^{-1}$, and $\Theta=-5.5$ K. 

\begin{figure} [b]
\includegraphics[width=8.5cm]{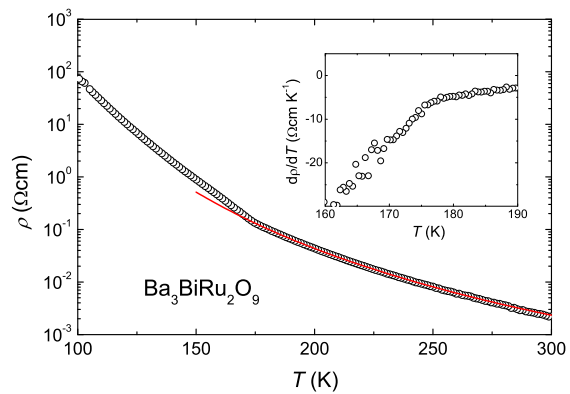}%
\caption{\label{fig4}(Color online) Electrical resistivity on a semilog scale. The solid line is a fit to Eq. \ref{eq.4}. The inset shows the temperature derivative of $\rho$ and its drastic change at $T^*=176$ K.}
\end{figure}

Fig. \ref{fig4} shows the temperature dependence of electrical resistivity $\rho$. Ba$_3$BiRu$_2$O$_9$ is nonmetallic - it exhibits increasing resistance with decreasing temperature - like other Ru 6H-type perovskites \cite{B105134M,Shlyk200764}. We tested various models for insulators, but the best fit was obtained with the variable-range hopping expression\cite{Mott:233761}:

\begin{equation}
\log\rho=a+bT^{\frac{-1}{1+n}}
\label{eq.4}
\end{equation}

where $n$ is the dimensionality of hopping and $a$ and $b$ are constants. This transport mechanism is expected for an insulating sample with strong inhomogeneity, where charge carriers are localised into states with various energies within a band gap. Transport between such states is realized with the help of a phonon, so that conduction occurs at a finite temperature. A least-square fit yields $a=-5.2$ $\Omega$cm, $b=99$ K$^{0.5}\Omega$cm and $n=0.97$ (\textit{i.e.} $\sim1$), pointing to 1D hopping. Below $T^*$, $\rho$ increases rapidly, seen as a drastic change in its temperature derivative (inset in Fig. \ref{fig4}), and Eq. \ref{eq.4} is no longer valid. This is reminiscent of superconducting cuprates, where out-of-plane resistivity increases due to the opening of a spin-gap \cite{PhysRevB.50.6534}.

\begin{figure} [t]
\includegraphics[width=8.5cm]{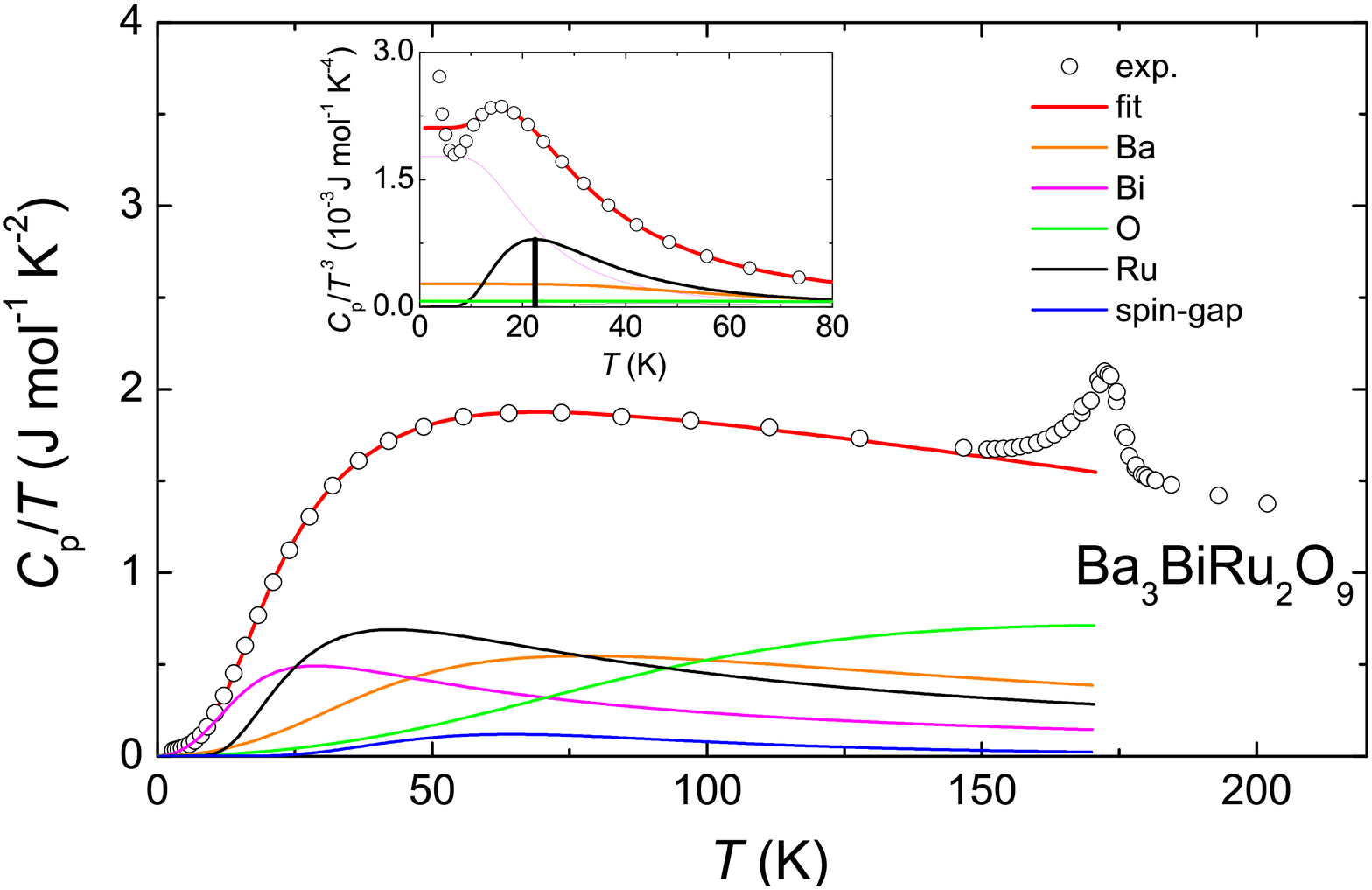}%
\caption{\label{fig5}(Color online) Temperature dependencies of the $C_{\rm{p}}/T$ ratio with its deconvolution into phononic and magnetic contributions (for details see the text). The inset shows apparent Einstein modes with $\Theta_E^{\rm{Ru}}=111$ K.}
\end{figure}

Fig. \ref{fig5} shows the temperature dependence of the heat capacity $C_{\rm{p}}/T$ ratio. The main feature is a sharp anomaly at $T^*$, suggesting a large release of magnetic and/or structural entropy. The total heat capacity of a solid consists of lattice, magnetic, and electronic contributions, but without a suitable phononic reference material (isostructural but with no unpaired $d$ electrons) the deconvolution of $C_{\rm{p}}$ is nontrivial. Initial attempts using a lattice term in the Debye approximation with a single characteristic frequency failed to produce a satisfactory fit. We have therefore used the approach of Junod \textit{et al.} \cite{Junod1, PhysRevB.27.1568}, which has been successfully applied to a number of oxide and intermetallic systems \cite{PhysRevB.76.014523,PhysRevLett.80.4903,PhysRevB.78.064516}, where the lattice contribution to the heat capacity is described by a sum of independent Debye and Einstein terms assigned to different groups of atoms. After testing all combinations of Debye and/or Einstein modes assigned to different sublattices, the best $r^2$ coefficient was obtained by fitting $C_{\rm{p}}$ below 150 K to a combination of 3 Debye modes assigned to the Ba, Bi and O sublattices, one Einstein mode from the Ru sublattice, and an additional term describing local magnetic excitations across the gap $\Delta_{\rm{C}}$\cite{PhysRevB.72.214434}:

\begin{equation}
C_{\rm{sg}}=nR\frac{(\Delta_{\rm{C}}/k_{\rm{B}}T)^2\exp(-\Delta_{\rm{C}}/k_{\rm{B}}T)} {1+n\exp(-\Delta_{\rm{C}}/k_{\rm{B}}T)}
\label{eq.5}
\end{equation}

where $n=3$ for singlet-triplet excitations and $R$ is the gas constant. We neglected the electronic contribution because the sample was found to be semiconducting. The best fit yielded $\Theta_{\rm{D}}^{\rm{Bi}}=101(1)$, $\Theta_{\rm{D}}^{\rm{Ba}}=296(4)$, $\Theta_{\rm{D}}^{\rm{O}}=630(4)$, $\Theta_{\rm{E}}^{\rm{Ru}}=111(1)$, and $\Delta_C/k_{\rm{B}}=193(6)$ K. Note that $\Delta_{\rm{C}}$, estimated using this methodology, is of the same order as $\Delta_{\chi}$. These contributions are shown in Fig. \ref{fig5}. The inset shows the $C_{\rm{p}}/T^3$ ratio, where the maximum at 16 K is an apparent hallmark of Einstein modes \cite{Junod1, PhysRevB.27.1568}. Neither a Debye model nor the spin-gap formula were able to reproduce this feature. The mean spin-gap from susceptibility and heat capacity $\overline{\Delta}=(\Delta_{\chi}+\Delta_{C}){/}{2k_{\rm{B}}}\approx 220$ K is equal to $0.7J_0{/}{k_{\rm{B}}}$. This is in perfect agreement with the ${\Delta}{/}{J_0}$ ratio calculated for Ba$_3$Mn$_2$O$_8$ ($\Delta{/}{k_{\rm{B}}}=12.2$ K and $J_0{/}{k_{\rm{B}}}=$17.4 K\cite{PhysRevB.72.214434}), which has a similar structure consisting of hexagonal layers of $S=1$ Mn$^{5+}$ dimers. \cite{PhysRevB.72.214434}. 

The transition at $T^*$ must be related to these changes in inter- and intradimer exchange interactions. We therefore performed \textit{ab initio} calculations on supercells for the structures refined at 300 and 2 K . As differences between total energies of different spin orientations appear due to the spin degrees of freedom, one can map total energies onto the Heisenberg hamiltonian:

\begin{equation}
\widehat{H} = -\sum_{<ij>}J_{ij}\overrightarrow{S_{i}}\overrightarrow{S_{j}}
\end{equation}

with one intradimer exchange parameter $J_0$ and two interdimer ones, $J_{ab}=(4J_1+2J_2)/6$ (in-plane) and $J_c=(J_3+J_4+J_5)/3$ (along $c$) -please refer to the inset in fig. \ref{fig3}. At 300 K, intradimer coupling was estimated as $J_0^{\rm{HT}}/k_{\rm{B}}=259$ K in the presence of interdimer couplings $J_{ab}^{\rm{HT}}/k_{\rm{B}}=-2.4$ K and $J_{c}^{\rm{LT}}/k_{\rm{B}}=5.4$ K, pointing to the domination of exchange within Ru$_2$O$_9$ units and its AFM character. At 2 K, the parameters are $J_0^{\rm{LT}}/k_{\rm{B}}=216$ K, $J_{ab}^{\rm{LT}}/k_{\rm{B}}=5.7$ K and $J_{ab}^{\rm{LT}}/k_{\rm{B}}=9.2$ K. These values of $J_0$ are somewhat smaller than obtained from susceptibility ($J_0/k_{\rm{B}}=320$ K), but significant contributions from next-nearest neighbors ($J_{ab}$ and $J_{c}$) justify use of the interacting dimer model. It is worth noting that both AFM exchange integrals $J_{ab}$ and $J_{c}$ at 2 K suggest magnetic frustration, which was proposed as the origin of the spin-gap opening in Mo$_3$Sb$_7$ \cite{PhysRevLett.100.137004,PhysRevLett.101.126404}.

In conclusion, we observe a spin-gap opening in the $S=1$ system Ba$_3$BiRu$_2$O$_9$ below $T^*=$176 K, with a spin-gap value of $\sim{220}$ K. A magnetoelastic effect is observed as a decrease in the Ru$-$Ru distance within Ru$_2$O$_9$ dimers below $T^*$, as well as in the magnetic, thermodynamic and electronic properties. An increase of unit cell volume is tentatively associated with a relaxation of the structure at $T^*$, although we cannot exclude other possibilities such as orbital ordering. \textit{Ab initio} calculations confirm strong AFM coupling within Ru$_2$O$_9$ dimers. Calculations to investigate the strong correlation effects typical of low-dimensional systems, using +\textit{Hubbard U} methods, are now required, as well as direct confirmation of the spin-gapped state by inelastic neutron scattering.

\acknowledgements{This work was supported by the ARC (DP0984585, DP0877695), AINSE, and the AMRFP. Dr Kevin Knight of ISIS assisted in HRPD data collection.}

\bibliography{Ba3BiRu2O9}

\end{document}